\documentclass[aps, prd, twocolumn, nofootinbib, floatfix, superscriptaddress]{revtex4-1}
\usepackage{comment}
\usepackage{graphicx}
\usepackage{dcolumn}
\usepackage{bm}
\usepackage{subfigure}
\usepackage{epsfig}
\usepackage{amsmath}
\usepackage{amsfonts}
\usepackage{graphicx}
\usepackage{amssymb}
\usepackage{nccmath}
\usepackage{amssymb,amsmath,amsfonts}
\usepackage{xfrac}
\usepackage{color}
\usepackage{tikz}
\usepackage[T1]{fontenc}
\usepackage[utf8]{inputenc}
\usepackage{ulem}

\usepackage{tcolorbox}
\usepackage{hyperref}
\usepackage{booktabs}
\hypersetup{colorlinks=true, citecolor=red, linkcolor=blue, urlcolor=blue}

\definecolor{rossos}{cmyk}{0,1,1,0.55}
\definecolor{bluscuro}{rgb}{0.15, 0.2, .85}
\definecolor{bluchiaro}{cmyk}{1,.3,0.,0.1}

\let\oldsqrt\sqrt
\def\sqrt{\mathpalette\DHLhksqrt}
\def\DHLhksqrt#1#2{%
\setbox0=\hbox{$#1\oldsqrt{#2\,}$}\dimen0=\ht0
\advance\dimen0-0.2\ht0
\setbox2=\hbox{\vrule height\ht0 depth -\dimen0}%
{\box0\lower0.4pt\box2}}

\newcommand{\sss}[1]{{\scriptscriptstyle{#1}}}

\newcommand{\uPl}{\mathrm{Pl}}

\newcommand{\usssPl}{\sss{\uPl}}

\newcommand{\Mp}{M_\usssPl}

\newcommand{\beq}{\begin{equation}}
\newcommand{\eeq}{\end{equation}}
\newcommand{\bea}{\begin{equation}\begin{aligned}}
\newcommand{\eea}{\end{aligned}\end{equation}}

\newlength{\wsingfig}
\newlength{\wdblefig}
\newlength{\wquadfig}
\newlength{\wtriplefig}
\newcommand{\Eq}[1]{Eq.~(\ref{#1})}

\newcommand{\Fig}[1]{Fig.~{\ref{#1}}}

\newcommand{\be}{\begin{equation}}

\def\mnras{MNRAS}

\begin{document}
\rightline{}
\title{Primordial black holes as cosmic expansion accelerators}

\author{Konstantinos Dialektopoulos}
\email{kdialekt@gmail.com}
\affiliation{Department of Mathematics and Computer Science,
Transilvania University of Brasov, 500091, Brasov, Romania}

\author{Theodoros Papanikolaou}
\email{t.papanikolaou@ssmeridionale.it}
\affiliation{Scuola Superiore Meridionale, Largo San Marcellino 10, 80138 Napoli, Italy}
\affiliation{Istituto Nazionale di Fisica Nucleare (INFN), Sezione di Napoli, Via Cinthia 21, 80126 Napoli, Italy}

\author{Vasilios Zarikas}
\email{vzarikas@uth.gr}
\affiliation{Department of Mathematics, University of Thessaly, 35100, Lamia, Greece}



\begin{abstract}

We propose a novel and natural mechanism for cosmic acceleration driven by primordial black holes (PBHs) exhibiting repulsive behavior. Using a new ``Swiss Cheese'' cosmological approach, we demonstrate that this cosmic acceleration mechanism is a general phenomenon by examining three regular black hole spacetimes - namely the Hayward, the Bardeen and the Dymnikova spacetimes - as well as the singular de Sitter-Schwarzschild spacetime. 
Interestingly, by matching these black hole spacetimes with an isotropic and homogeneous expanding Universe, we obtain a phase of cosmic acceleration that ends at an energy scale characteristic to the black hole parameters or due to black hole evaporation. This cosmic acceleration mechanism can be relevant either to an inflationary phase with a graceful exit and reheating or to an early dark energy type of contribution pertinent to the Hubble tension. 
Remarkably, we find that ultra-light PBHs with masses $m<5\times 10^8\mathrm{g}$ dominating the energy content of the Univese before Big Bang Nucleosynthesis, can drive a successful inflationary expansion era without the use of an inflaton field. Additionally, PBHs with masses $m \sim 10^{12}\mathrm{g}$ and abundances $0.107 < \Omega^\mathrm{eq}_\mathrm{PBH}< 0.5$, slightly before matter-radiation equality, can produce a substantial amount of early dark energy, helping to alleviate the $H_0$ tension. 
\end{abstract}


\maketitle

\medskip
{\bf{Introduction}}--
Primordial black holes (PBHs), were firstly introduced in `70s~\cite{1967SvA....10..602Z, Carr:1974nx,1975ApJ...201....1C,1979A&A....80..104N}, leading to the famous Hawking discovery of black hole evaporation~\cite{Hawking:1975vcx}. Among numerous PBH formation mechanisms one can mention indicatively the collapse of enhanced primordial cosmological perturbations~\cite{Carr:1974nx,Carr:1975qj,Musco:2020jjb} [See~\cite{Garcia-Bellido:1996mdl,Yokoyama:1995ex,Garcia-Bellido:2017mdw} for inflationary realizations], cosmological phase transitions~\cite{Hawking:1982ga,Moss:1994iq,Jung:2021mku}, scalar field instabilities~\cite{Khlopov:1985fch}, collisions of highly energetic particles~\cite{Saini:2017tsz}, topological defects~\cite{Hawking:1987bn,Polnarev:1988dh} and modified/quantum gravity constructions~\cite{Barrow:1996jk,Kawai:2021edk,Papanikolaou:2023crz}. 

Interestingly, these objects have recently captured the attention of the scientific community since they can potentially account for a part or the totality of dark matter~\cite{Chapline:1975ojl,Carr:2020xqk}, providing us with a natural explanation for the formation of large-scale structures~\cite{Meszaros:1975ef,Afshordi:2003zb}. Furthermore, PBHs can interpret quite well some of the black hole merging events recently detected by the LIGO/VIRGO collaboration~\cite{Sasaki:2016jop,Franciolini:2021tla}, playing also a significant role in the processes of reheating~\cite{Lennon:2017tqq}, baryogenesis~\cite{Barrow:1990he, Baumann:2007yr, Aliferis:2014ofa, Aliferis:2020dxr} and cosmic magnetic field generation~\cite{Safarzadeh:2017mdy,Araya:2020tds,Papanikolaou:2023nkx}. For recent reviews on PBHs, see~\cite{Carr:2020gox,Escriva:2022duf}.

To the best of our knowledge, most research on PBHs assumes Schwarzschild or Kerr spacetimes for the PBH metric~\cite{Khlopov:2008qy,Carr:2016drx,Escriva:2022duf,LISACosmologyWorkingGroup:2023njw,Choudhury:2024aji}. In this Letter, however, we explore PBHs with repulsive behaviour, which, in most cases, imply that they are regular~\cite{Luongo:2023aib}. These classes of PBH spacetimes are particularly interesting from a theoretical perspective, as they offer a potential common solution to both the dark matter and singularity problems~\cite{Easson:2002tg,Dymnikova:2015yma,Pacheco:2018mvs,Arbey:2021mbl,Arbey:2022mcd,Banerjee:2024sao,Davies:2024ysj,Calza:2024fzo,Calza:2024xdh}. For comprehensive reviews on regular black hole spacetimes see~\cite{Ansoldi:2008jw,Nicolini:2008aj,Torres:2022twv,Lan:2023cvz}. 

With this Letter we aim to report a novel and natural mechanism that generates cosmic acceleration based on a ``Swiss Cheese'' approach, used to model a black hole dominated Universe. The ``Swiss Cheese'' cosmological model, firstly introduced in the '40s by Einstein and Strauss~\cite{Einstein:1945id}, matches the metric of a black hole with a homogeneous and isotropic spacetime. Notably, in the case of the Schwarzschild spacetime, 
one gets naturally a dust filled homogeneous and isotropic Universe. 

As we will show below, when the matching involves black holes with repulsive behavior, it generically leads to an era of accelerated cosmic expansion. To describe and analyze this cosmic acceleration mechanism, we will specifically consider three classes of regular black hole  spacetimes, namely the Hayward~\cite{Hayward:2005gi}, the Bardeen~\cite{1968qtr..conf...87B} and the Dymnikova~\cite{Dymnikova:1992ux,Dymnikova:2004qg, Dymnikova:1992ux,1996IJMPD...5..529D} spacetimes, presenting a repulsive de-Sitter like core, as well as one singular case, the de-Sitter Schwarzschild black hole~\cite{Kottler:1918cxc, Shankaranarayanan:2003ya}, which is repulsive at large distances.

Remarkably, we find that this cosmic acceleration mechanism can play a significant role in generating an inflationary expansion phase being terminated below a specific energy scale, depending on the PBH metric considered, or naturally due to black hole evaporation. 
Additionally, it may contribute to a substantial early dark energy (EDE) component, without the need of additional scalar fields, before the emission of the cosmic microwave background (CMB), thereby helping to alleviate the Hubble tension. 

\newpage
{\bf{Regular black holes}}--
Regular black hole solutions can arise naturally within quantum gravity scenarios, with the latter removing curvature singularities, working either with an ensemble of continuous quantum spacetimes \cite{Cadoni:2023nrm, Bonanno:2023rzk, dePaulaNetto:2023cjw} or with discrete  geometries~\cite{Singh:2009mz,Ashtekar:2023cod}. In both cases, the quantum gravity theory under consideration should explain how classical spacetime emerges. It is then reasonable to describe the black hole spacetime with a classical metric presenting phenomenological corrections at high curvature, so as that it exhibits non-singular (regular) behavior at the center~\cite{1968qtr..conf...87B,Frolov:1981mz,Roman:1983zza,Dymnikova:1992ux,Borde:1996df,Hayward:2005gi,Hossenfelder:2009fc,Bambi:2013gva,Bambi:2013caa, Hawking:2014tga, Frolov:2014jva,Bardeen:2014uaa,Haggard:2014rza, Barrau:2015uca,Haggard:2015iya}. 
In particular, within this study, we will work with three types of non-singular black hole spacetimes, namely the Hayward~\cite{Hayward:2005gi}, the Bardeen~\cite{1968qtr..conf...87B} and the Dymnikova~\cite{Dymnikova:1992ux,Dymnikova:2004qg} spacetimes. In the following, we illustrate our cosmic acceleration mechanism with the Hayward metric.

The metric originally proposed by Hayward reads as~\cite{Hayward:2005gi} 
\begin{equation}\label{eq:Hayward_metric}
\mathrm{d}s^2=-F(R)\,\mathrm{d}t^2+\frac{1}{F(R)}\,\mathrm{d}R^2+R^2\,\mathrm{d}\Omega^2 ~,
\end{equation}
with
\begin{equation}\label{hayward}
F(R)=1-\frac{2\,G_\mathrm{N} M(R)}{R}~,
\end{equation}
and
\begin{align}
M(R)&=\frac{m\,R^3}{R^3+2\,G_\mathrm{N}\, m\,L^2} \nonumber \\ 
\nonumber &\approx
\begin{cases}
m & (R  \gg G_\mathrm{N}^{1/3} m^{1/3} L^{2/3}) \\
R^3/(2G_\mathrm{N} L^2) & (R  \ll G_\mathrm{N}^{1/3} m^{1/3} L^{2/3})  \, .
\end{cases}
\end{align}
Here, we set $c=1$ and $m$ is interpreted as the mass of the black hole at asymptotic infinity.


This metric \eqref{eq:Hayward_metric} possesses Killing horizons, 
determined by the roots of 
\begin{equation} \label{hor}
F(R) = 1-\frac{2\,G_\mathrm{N}\,m\,R^2}{R^3+2\,L^2\,G_\mathrm{N}\, m}=0 \ , 
\end{equation}
 and dependent on 
$L$ and  $m$.  In particular, one can show that for $m>\frac{3\sqrt{3}L}{4G_\mathrm{N}}$the metric features inner and outer horizons at
$R \simeq 2\,G_\mathrm{N}\,m$ and $R \simeq L$, respectively. 
A detailed analysis of these Killing horizons including Penrose diagrams can be found in \cite{Hayward:2005gi,DeLorenzo:2014pta}. For the cases of the Bardeen and Dymnikova black hole spacetimes, see the discussion in the Appendices \ref{Appendix_Bardeen} and \ref{Appendix_Dymnikova} respectively. In the Appendix \ref{dSS}, we study also the case of a singular black hole spacetime with a repulsive behaviour at large distances, namely the de Sitter-Schwarzschild spacetime~\footnote{We need to note here that the de Sitter-Schwarzschild metric can be viewed as a particular limit of the McVittie metric~\cite{1933MNRAS..93..325M} describing a black hole or massive compact object within general relativity immersed in an expanding cosmological spacetime. According to recent arguments~\cite{DeLuca:2020jug,Hutsi:2021vha,Hutsi:2021nvs} it can describe quite well the PBH metric at least in the early stages of the PBH gravitational collapse.}.

{\bf{Cosmological ``Swiss Cheese'' matching gives acceleration}}--
The matching of an exterior homogeneous and isotropic spacetime to an arbitrary, static and spherically symmetric interior is described within the Appendix \ref{swch}. See also here~\cite{Kofinas:2017gfv}, \cite{Anagnostopoulos:2022pxa}, for a detailed discussion about matching conditions. At this point, it is important to clarify two issues regarding the ``Swiss Cheese'' approach adopted here. Firstly, in our case the ``Swiss Cheese'' concerns the early Universe, where an almost homogeneous distribution of black holes is a very good approximation~\cite{Desjacques:2018wuu, Ali-Haimoud:2018dau, MoradinezhadDizgah:2019wjf,DeLuca:2022uvz}, something that is not true for the present epoch. Secondly, the arguments against the stability of the ``Swiss Cheese'' model~\cite{Krasinski:1997yxj} are not important in our case since we do not use the matching to describe the influence of a current large scale structure, being the original idea of the Einstein-Strauss work. We study instead the effect on the early cosmic expansion considering a Universe filled with many black holes, hence our ``Swiss Cheese'' method is well justified.

For the case of the Hayward metric, our ``Swiss Cheese'' describing a single black hole matched to a homogeneous and isotropic Universe, gives rise the following cosmic expansion equations
\begin{gather}
H^2=\frac{\dot{a}^2}{a^2}=\frac{2G_\mathrm{N} m}{R^3+2 G_\mathrm{N} m L^2}-\frac{k}{a^2} \,, 
\label{eq:H_2_Hayward}\\
\frac{\ddot{a}}{a}=\frac{G_\mathrm{N} m (4\,G_\mathrm{N} L^2 m-R^3)}{(2 G_\mathrm{N} L^2 m+R^3)^2}, 
\label{acceleration eq real}
\end{gather}
where $R = a\,r_\mathrm{\Sigma}$ and $a$ and $H\equiv \dot{a}/a$ stand for the scale factor and the Hubble parameter respectively. The radial distance $r_\mathrm{\Sigma}$ called Schucking radius~\cite{Schucking:1954} and is the comoving distance in the cosmological spacetime frame at we which we perform the matching. For more details on the derivation of the aforementioned equations see the Appendix \ref{swch}.

A Universe filled with many black holes, the ``Swiss Cheese", arises by simply repeating the above process for many disjoint spherical regions in the FLRW background \cite{Einstein:1945id}, \cite{Schucking:1954}, \cite{Carrera:2008pi}. In particular, one excises multiple non-overlapping spheres from the FLRW spacetime, replaces each with a spherical black hole region and then matches it using the same junction conditions. As long as the excised spheres do not overlap, each black hole region is matched locally, and the remaining cosmological background still satisfies the gravitational field equations with the same matter content. This repetition is possible because field equations are local: each excised region can be matched independently. The large-scale homogeneity and isotropy is recovered because the holes can be distributed in a statistically homogeneous and isotropic way. 
Thus, one then needs to work with energy densities~\cite{Schucking:1954}, \cite{Carrera:2008pi}.  More specifically, every BH vacuole must have the same mass density with the mass density of the rest of the Universe, i.e. $\rho=3m/(4\pi\,(a\,r_{\Sigma})^3)$. Using the latter expression and solving for $m$ we can get from \Eq{acceleration eq real}, \Eq{acceleration with density}. For a more detailed explanation see \cite{Schucking:1954}, \cite{Carrera:2008pi}. See also Appendix \ref{swch} for more technical details.

One then can determine the conditions for cosmic acceleration or deceleration in such a Universe, writing
\begin{equation}
\frac{\ddot{a}}{a}= 4 \pi G_\mathrm{N} \rho \,  \frac{(16 G_\mathrm{N} L^2 \pi \rho-3)}{(8 G_\mathrm{N} L^2 \pi \rho+3)^2} \, .
\label{acceleration with density}
\end{equation}
The expression $16 G_\mathrm{N} L^2\,\pi\,\rho-3$ in the numerator, determines the sign of the acceleration parameter $\ddot{a}/a$ with the characteristic density, $\rho_{\rm ch}$, at which 
 $\ddot{a}=0$, reading 
\begin{equation}\label{rhoch}
\rho_{\mathrm {ch}}=\frac{3}{16 \pi G_\mathrm{N}L^2} \, .
\end{equation}
When $\rho>\rho_{\mathrm {ch}}$, the acceleration is positive
whereas one recovers a conventional deceleration when $\rho<\rho_{\mathrm {ch}}$. Consequently, the exit from inflation is natural and guaranteed without introducing any
fine tuning.


{\bf{Cosmological implications of the PBH driven cosmic acceleration mechanism}}-- Having derived the modified cosmic expansion dynamics for a Universe filled with PBHs with repulsive behavior, we now explore two possible cosmological implications of this PBH-driven cosmic acceleration: its role in generating an early inflationary cosmic era as well as its possibility to account for an early dark energy component. 

{\textit{Inflation with graceful exit and reheating}}--
Since the initial conditions concerning the Big Bang are unknown and perhaps unknowable, a black hole dominated early Universe remains a plausible possibility~\cite{Nagatani:1998gv,Conzinu:2023fui}. We thus assume an initial population of randomly distributed Hayward PBHs, dominating the energy budget of the Universe before Big Bang Nucleosynthesis (BBN), formed in the early Universe, usually due to quantum gravity gravitational collapse processes~\cite{1968qtr..conf...87B,Bojowald:2005qw,DeLorenzo:2014pta, Bonanno:2020fgp,Shafiee:2022jfx,Carballo-Rubio:2023mvr,Bambi:2023try,Bonanno:2023rzk,Harada:2025cwd}. 
Even if the early Universe initially consisted of a cosmic soup of massless degrees of freedom within the framework of Standard Model Hot Bing Bang cosmology~\cite{Kolb:1990vq}, together with PBHs, PBHs will eventually dominate due to their slower dilution compared to radiation. For a more detailed discussion on how PBHs can outlast and dominate radiation, see~\cite{Hidalgo:2011fj,Suyama:2014vga,Zagorac:2019ekv,Hooper:2019gtx}.    
Working within flat topology ($k=0$), the expansion rate \eqref{eq:H_2_Hayward} will read as
\begin{equation}
H^2=\frac{8\,\pi}{3}\,G_\mathrm{N}\,\left( \rho^{-1} +\frac{8\pi}{3}\,G_\mathrm{N}\,L^2\right) ^{-1} \, .
\label{modH^2}
\end{equation}
The solution of this equation provides a non-singular expansion, as it is expected since our ``Swiss Cheese'' Universe is
filled with non-singular black holes.
For a very large energy densities $\rho$, \Eq{modH^2} becomes $H^2\simeq L^{-2}=\mathrm{constant}$, giving rise to a nearly exponential expansion phase with the scale factor reading as
\begin{equation}
    a\simeq a_\mathrm{i}e^ {t/L}, 
\end{equation}
where $a_{\rm i}$ stands for a non-zero initial scale factor. Inflation can end either due to black hole evaporation or when the black hole density reaches $\rho_{\rm e} \simeq G_\mathrm{N}^{-1}L^{-2}$. This occurs when the two terms in \Eq{modH^2} become comparable, indicating the end of inflation, as denoted by the subscript $e$. Thus, after the end of inflation, if evaporation has not yet occurred, the expansion rate can be expressed as 
\beq
H^2=\frac{8\,\pi}{3}\,G_\mathrm{N}\, \frac{\rho_{\rm e}\,a_{\rm e}^3} {a^3} \, , 
\eeq
since the second term in \Eq{modH^2} becomes negligible. Here, we have used the fact that energy density of the Universe is equal to the PBH mass density, i.e. $\rho=\rho_\mathrm{PBH}=\rho_{\rm e} a^3_{\rm e}/a^3$. 

It is a reasonable approximation to assume that PBHs evaporate rapidly~\cite{Hawking:1974rv}, with  their lifetime depending on their initial mass. This evaporation process can take place either after the end of inflation or during the inflationary era, leading to an exit from inflation before reaching the critical density $\rho_{\rm e} \simeq G_\mathrm{N}^{-1} L^{-2}$. After black hole evaporation, the expansion rate follows that of a radiation-dominated Universe, with $H^2\propto a^{-4}$. 

Let us then make a qualitative analysis regarding the number of e-folds during inflation $N_\mathrm{inf}$, defined as $N_\mathrm{inf}\equiv\ln\left( a_{\rm e}/a_{\rm i}\right),$ as well as reheating within our scenario. We work for concreteness with the case of evaporation taking place after the end of inflation. As we will show later this is always the case for regular PBHs independently of the the values of the PBH mass $m$ and the PBH regularising parameters. One then gets from \Eq{modH^2} that
\beq
    \rho_{\rm e} \simeq G_\mathrm{N}^{-1} L^{-2}=\frac{\pi^2}{30}\,g_{\rm reh}\,T_{\rm reh}^4\,,
\eeq
where $g_{\rm reh}$ is the number of relativistic degrees of freedom of the primordial plasma and where we have assumed instantaneous reheating. The reheating temperature $T_\mathrm{reh}$ can then be recast as
\begin{equation}
T_\mathrm{reh}=\left( \frac{30}{\pi^2\,g_\mathrm{reh}}\right) ^{1/4} L^{-1/2}G_\mathrm{N}^{-1/4}\,. 
\end{equation}
Since the temperature of the thermal plasma $T$ scales as $T\propto 1/a$, we have that $a_{\mathrm{e}}/ a_0 \sim T_0/T_{\rm reh} $ with $T_0=2.7K$ and $a_0=1$, where $0$ denotes our present epoch. 

One then can derive a lower bound on $N_\mathrm{inf}$ in order to solve the horizon problem. In particular, one should require that the today's comoving observable length scale, $\lambda_0$, defined as $ \lambda_0\equiv  (a_0\,H_{0})^{-1}$, is smaller than the particle 
horizon distance during inflation $d_\mathrm{inf}$, defined as 
$d_\mathrm{inf}\equiv 1/(a_\mathrm{i}H_{\rm{inf}})$. Consequently, one has that 
\beq\label{eq:N_inf_constraint}
\begin{split}
\lambda_0/d_\mathrm{inf}<1 & \Leftrightarrow  e^{-N_\mathrm{inf}}a_{\rm e}\,H_{\rm{inf}}/(a_{0}\,H_{0})<1  \\ & \Leftrightarrow 
N_\mathrm{inf}>\ln (T_0/T_{\mathrm{reh}})-\ln (L\,H_0).
\end{split}
\eeq
This bound on  $N_\mathrm{inf}$ is always satisfied for reasonable values of $L$ above the Planck length and small enough $a_{\rm i}$.

Reheating is also vital for any viable inflationary scenario. In our scenario, it happens naturally due to the evaporation of PBHs after the end of the exponential inflationary phase. For a crude estimate of the PBH evaporation timescale, we use a time-dependent version of the Hayward metric
giving rise to an improved evaporation law reading as~\cite{Frolov:2017rjz}\footnote{One should note here that evaporation happens very fast, almost instantaneously, compared to other relevant time scales~\cite{Calza:2024fzo,Calza:2024xdh}. This allows us to make the common approximation that up to the last stages of the evaporation process we have a ``Swiss Cheese" PBH-dominated Universe while at the very end of the PBH evaporation one is met with a mixture of matter and radiation necessitating the use of a Vaidya-type modification of the time-dependent Hayward metric in order to deduce the  improved evaporation law \Eq{evap}~\cite{Frolov:2016pav,Lorenzo:2024pisa,Frolov:2016gwl}.}
\begin{equation}\label{evap}
{dm(t)\over dt}\sim - {n_{\rm p}\over 1920\,\pi}{l_\mathrm{Pl}^2\over L^2}{1\over G_\mathrm{N}^2 \,m(t)^2} \sim {1\over C^3 G_\mathrm{N}^2 \,m(t)^2} \, ,
\end{equation}
where $n_{\rm p}$ is the number of distinct polarizations of emitted particles, $l_\mathrm{Pl}$ is the Planck length and  
$C \equiv  \left(\frac{640 \pi}{n_{\rm p}}\right)^{1/3}\left(\frac{L}{l_\mathrm{\rm Pl}}\right)^{2/3} $.
Integrating thus \Eq{evap} from $m$ to $0$, the black hole lifetime we get is
\begin{equation}\label{firstconstraint}
t_{\rm evap}= \frac{1}{3} \, C^3\,G_\mathrm{N}^{2}\,m^3 \, . 
\end{equation} 
For successful reheating we require as well that evaporation is complete before BBN, which begins at
$t_{\rm BBN} \simeq 1\mathrm{min}$,
\begin{equation}\label{constr1}
t_{\rm evap}\leq t_{\rm BBN} \Rightarrow  m \leq G_\mathrm{N}^{-2/3}\, C^{-1} (3t_{\rm BBN})^{1/3} \, .
\end{equation}
This gives us a constraint on the black holes mass. In particular, for $n_{\rm p}\sim 100$ and $L=100\,l_{\rm Pl}$, we get $m< 10^{32}{\rm GeV}=5 \times 10^8$g.

In order to be consistent with BBN, one should require that the thermalisation of
Hawking radiation after 
evaporation results in a temperature $T_{\rm reh}$ larger than the BBN, $T_{\rm BBN}\sim 4\mathrm{MeV}$, leading to 
\begin{equation}\label{constr2}
T_{\rm reh}>T_{\rm BBN} \Rightarrow  \rho_{\rm PBH,evap} > \frac{\pi^2\, g_{\rm reh}}{30}T_{\rm BBN}^4 \, .
\end{equation}
In summary, as long as our scenario satisfies the constraints \eqref{eq:N_inf_constraint}, \eqref{constr1} and \eqref{constr2}, it is consistent with an inflationary epoch with graceful exit and reheating proceeding through PBH evaporation.

\begin{figure*}[ht!]
\centering
\includegraphics[width=0.49\textwidth]{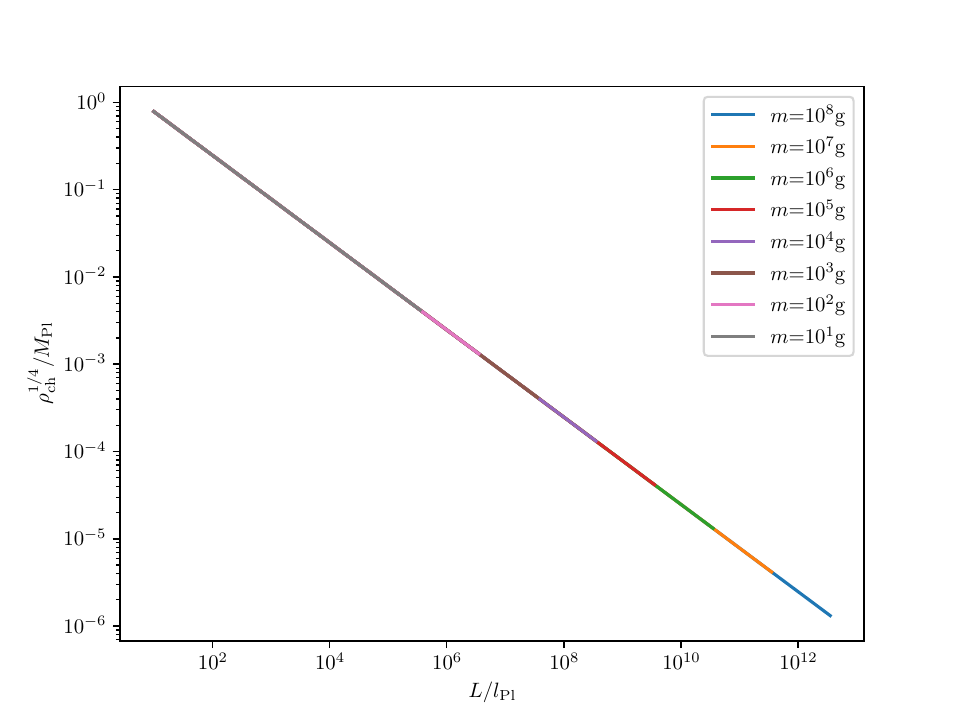}
\includegraphics[width=0.49\textwidth]{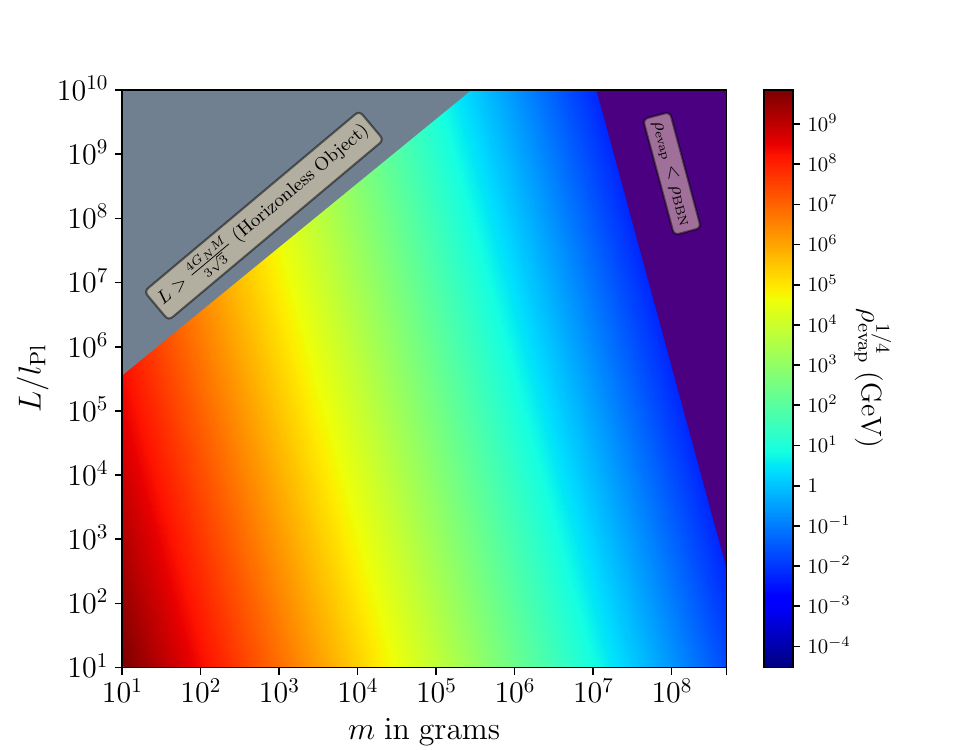}
\caption{{Left Panel: We show $\rho_\mathrm {ch}$ as a function of the regularising parameter $L$. $L$ is bounded from above, i.e. $L<\frac{4G_\mathrm{N}m}{3\sqrt{3}}$. Right Panel: We show $\rho_\mathrm{evap}$ (color-bar axis) as a function of the PBH mass $m$ ($x$-axis) and the regularising parameter $L$ ($y$-axis).The grey region $L>\frac{4G_\mathrm{N}m}{3\sqrt{3}}$ is not of interest since there one is met with horizonless objects while the black one is excluded since there $\rho_\mathrm{evap}<\rho_\mathrm{BBN}$.}}
\label{fig:rho_c_vs_rho_evap}
\end{figure*}

At this point, it is useful to compare the characteristic energy density $\rho_{\rm ch}$ \eqref{rhoch} and the energy density at the end  of the PBH evaporation process $\rho_\mathrm{evap}$ in order to see at which energy scale we are met with the termination of the early cosmic acceleration phase. One can estimate  $\rho_\mathrm{evap}$ by assuming having a radiation-dominated Universe with $a\propto t^{1/2}$ right after PBH evaporation. Thus, one has $\rho_\mathrm{evap} = \frac{3}{8\pi G_\mathrm{N}}H^2_\mathrm{evap}$, with $H_\mathrm{evap}=\frac{1}{2t_\mathrm{evap}}$, where $t_\mathrm{evap}$ is given by \Eq{firstconstraint}. After some simple algebra one can show that
\beq\label{eq:rho_evap}
\rho_\mathrm{evap} = 27648\pi^4\Mp^4\left(\frac{\Mp}{m}\right)^6\left(\frac{n_\mathrm{p}}{640\pi}\right)^{2/3}\left(\frac{l_\mathrm{Pl}}{L}\right)^{4/3},
\eeq
where we have used the fact that $\Mp^2 = \frac{1}{8\pi G_\mathrm{N}}$. In \Fig{fig:rho_c_vs_rho_evap} we plot in the left panel $\rho_{\rm ch}$ as a function of $L$ accounting for the upper bound on $L$, i.e. $L<\frac{4G_\mathrm{N}m}{3\sqrt{3}}$, so as to have a horizonfull object. As one may observe, $\rho_{\rm ch}$ is quite close to the Planck scale, with the smaller PBH masses giving a higher $\rho_{\rm ch}$ compared to the smaller ones. In the right panel of \Fig{fig:rho_c_vs_rho_evap} we show $\rho_\mathrm{evap}$ \eqref{eq:rho_evap} in the color bar axis as a function of $m$ in the $x$-axis and $L$ in the $y$-axis. The grey region $L>\frac{4G_\mathrm{N}m}{3\sqrt{3}}$ is not of particular interest since there one is met with horizonless objects while the black one is excluded since there PBH evaporation takes place after BBN, i.e. $\rho_\mathrm{evap}<\rho_\mathrm{BBN}$. The condition $\rho_\mathrm{evap}<\rho_\mathrm{BBN}$ is equivalent to 
\beq
L>7\times 10^{5} l_\mathrm{Pl}\left(\frac{10\mathrm{MeV}}{\rho^{1/4}_\mathrm{BBN}}\right)^3\left(\frac{n_\mathrm{p}}{640\pi}\right)^{1/2}\left(\frac{10^8\mathrm{g}}{m}\right)^{9/2}.
\eeq

As one can infer from both panels $\rho_\mathrm {ch}$ is always larger than $\rho_\mathrm{evap}$, meaning that the cosmic acceleration phase is always terminated before PBH evaporation. Similar conclusions can be drawn as well for the other two regular PBH spacetimes considered in the appendices since one in general finds prolonged regular black hole lifetimes with their evaporation time being quite well approximated as $t_\mathrm{evap}\propto \tilde{C} t_\mathrm{S}$, where $t_\mathrm{S}$ is the  Schwarzschild black hole evaporation time and $\tilde{C}$ is a parameter depending on the regularising PBH metric parameters and which is greater than unity, i.e. $\tilde{C}>1$~\cite{Calza:2024fzo}. For the Dymnikova case, \Fig{fig:rho_c_vs_rho_evap} is more complicated since there is no analytic expression for $\rho_\mathrm {ch}$ as a function of the regularising parameters. With regard to the singular Schwarzschild de-Sitter black hole spacetime, one finds that the early cosmic acceleration phase terminates always due to PBH evaporation [See Appendix \ref{dSS}.].

Up to now, we studied the background behaviour. It is important then to make here a comment regarding the generation of cosmological perturbations within our inflationary mechanism. Notably, in an inflationary mechanism, not driven by an
inflaton scalar field, like in our case, primordial curvature perturbations can be potentially generated by massive spectator scalar fields, becoming light during inflation, like the
curvaton one~\cite{Lyth:2001nq}. 
Recently, there was also reported a relevant mechanism, where primordial scalar/curvature perturbations are generated via
second-order tensor eﬀects arising from the instability of the dS space~\cite{Bertacca:2024zfb}. These two possibilities for the curvature perturbation generation can at least be applied to the cases of de Sitter-Schwarzschild and Dymnikova black holes driving inflation, being solutions of GR. 
On the other hand,
quantum/modified gravity theories beyond GR, accepting regular black hole solutions, like the Hayward and the Bardeen ones, are usually characterised by bouncing
cosmological setups, where nearly scale-invariant curvature power spectra can be
easily realised~\cite{Lilley:2015ksa, Battefeld:2014uga, 
Peter:2008qz}, being thus compatible with CMB observations~\cite{Cai:2014bea,Cai:2014xxa}.

A full perturbation analysis requires to specify a particular gravitational action and the respective field equations, an analys that is going beyond the scope of the current work, whose findings regarding a repulsive PBH driven inflationary mechanism 
are rather model independent.

{\textit{Early dark energy contribution before matter-radiation equality--}}
We consider now a different cosmic scenario, where inflation has already been occurred due to an inflaton field or any other possible mechanism. In such a case, if the initial PBH abundance is not sufficiently high, the PBH abundance will increase, being diluted slower compared to radiation, but not enough to dominate, i.e. $\Omega_\mathrm{PBH}<0.5$.  
One then faces after the end of inflation the co-existence of radiation and PBHs and an extended ``Swiss Cheese'' model that accounts for both PBHs and radiation within a spherical vacuum hole (or Szekeres type model) is needed~\cite{Carrera:2008pi,Celerier:2024dvs}.

Interestingly enough, if the evaporation process begins slightly before matter-radiation equality, corresponding to PBH masses of around $10^{12}\mathrm{g}$, with PBH abundances in the range $0.107<\Omega_\mathrm{PBH}<0.5$, one gets the correct amount of EDE consistent with current observational constraints on EDE from CMB~\cite{Pettorino:2013ia} and the growth of large scale structure formation~\cite{Smith:2020rxx} reading as 
\beq
\Omega_\mathrm{EDE}(t_\mathrm{LS})<0.015\quad\mathrm{and}\quad 0.015<\Omega_\mathrm{EDE}(t_\mathrm{eq})<0.107,
\eeq
where $t_\mathrm{LS}$ and $t_\mathrm{eq}$ stand for the times at the last-scattering and matter-radiation equality respectively.

Remarkably, in such a scenario we have an EDE component in form of PBHs without the addition of exotic scalar fields as usually adopted in the literature~\cite{Poulin:2023lkg}. An advantage of our mechanism is that our EDE type of contribution decays faster than radiation due to Hawking evaporation, a condition which is necessary to lead to an increased value of the Hubble parameter at early times. Therefore, it can be consistent with late-time SNIa measurements~\cite{Poulin:2018cxd,Kumar:2024soe, Shah:2023sna}, alleviating in this way naturally the $H_0$ cosmic tension.

{\bf{Conclusions}}--We propose in this Letter a novel and natural mechanism of cosmic acceleration triggered by a population of PBHs with repulsive behaviour, dominanating the early Universe and giving rise within the context “Swiss Cheese” cosmology to an inflationary de Sitter-like expansion.
In particular, by matching such repulsive black hole spacetimes with an isotropic and homogeneous metric, we always find a stage of an early cosmic acceleration that can end naturally below a characteristic energy scale depending on the regularisation parameters of the PBH metric considered or due to PBH evaporation.

We further show that this generic repulsive PBH driven cosmic acceleration mechanism can explain either inflation or an EDE injection before CMB emission. 
Notably, our analysis shows that ultra-light PBHs with masses $m<5\times 10^8\mathrm{g}$, having repulsive behaviour, can drive an exponential inflationary phase being compatible with a graceful exit and successful reheating proceeding through black hole evaporation. Furthermore, PBHs with masses  $m \sim 10^{12}\mathrm{g}$ and abundances $0.107 < \Omega^\mathrm{eq}_\mathrm{PBH}< 0.5$ slightly before matter-radiation equality can produce a substantial amount of EDE so as to alleviate the Hubble tension. 

\begin{figure}[h!]
\centering
\includegraphics[width=0.49\textwidth]{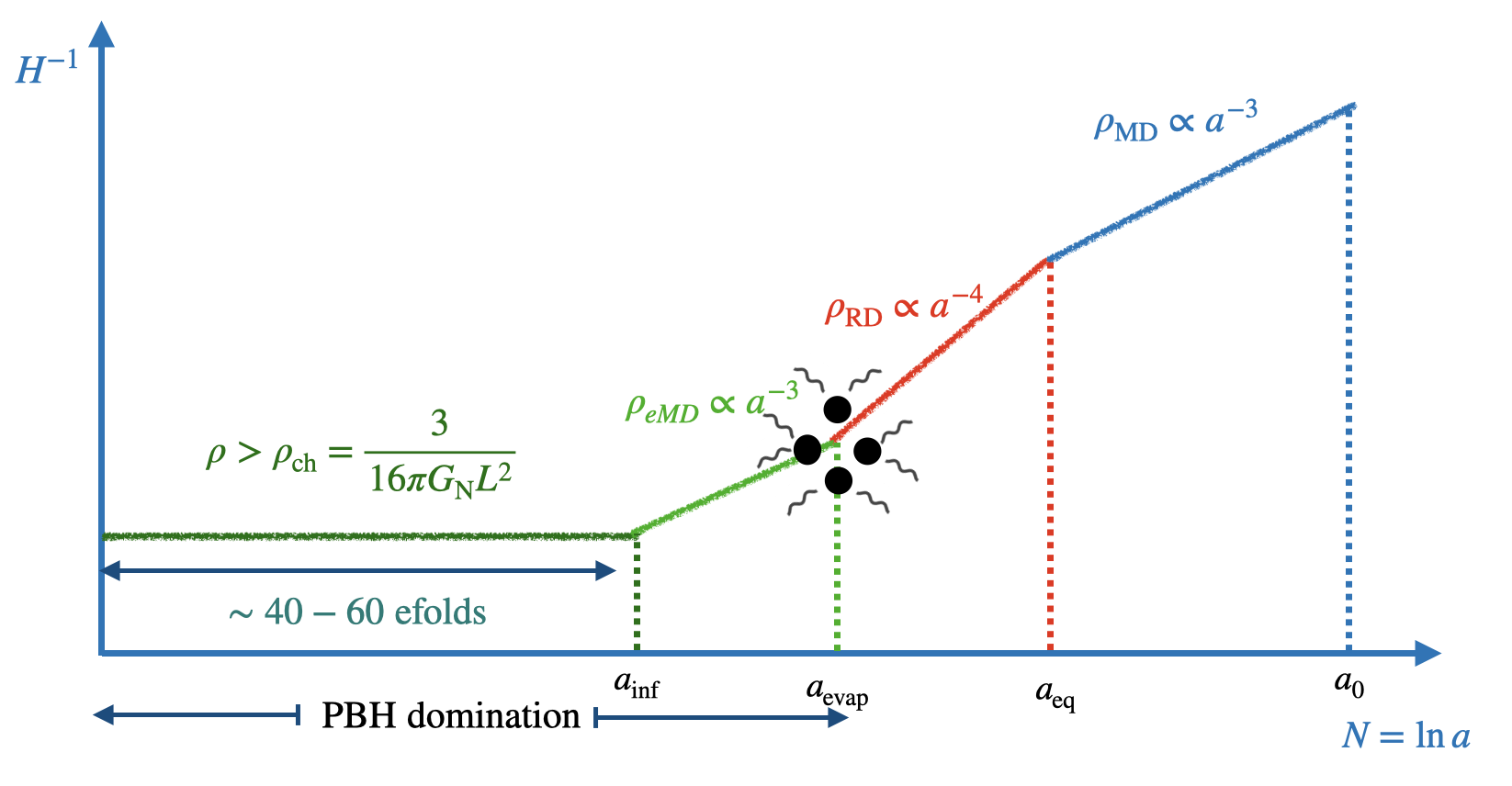}
\caption{{The evolution of the cosmological horizon $H^{-1}$ for a Universe filled with ``repulsive-like" primordial black holes of Hayward type.}}
\label{fig:cosmic_history}
\end{figure}

Our alternative cosmological scenario responsible for the generation of early cosmic acceleration is illustrated in the schematic \Fig{fig:cosmic_history} where we show the evolution of the cosmological horizon $H^{-1}$ as a function of the e-fold number $N$ for the case of the regular Hayward metric considered above. Similar figures are expected also for the Bardeen, Dymnikova and the Schwarzschild-de Sitter black holes considered in the Appendices \ref{Appendix_Bardeen}, \ref{Appendix_Dymnikova} and \ref{dSS}.

Interestingly enough, such repulsive black hole geometries can be naturally realized within GR. This is the case for the singular de Sitter-Schwarzschild~\cite{Kottler:1918cxc} and the regular Dymnikova geometries~\cite{Dymnikova:1992ux}. Furthermore, modified/quantum gravity theories can also accept repulsive black hole solutions, which are generically regular at the center \cite{Bambi:2023try}. In particular, the Hayward black hole geometry can be realized in asymptotic safe gravity~\cite{Bonanno:2020fgp} as well as within higher-dimension gravity theories~\cite{Bueno:2024dgm} while the Bardeen black hole is generically realised in gravity theories coupled to non-linear electrodynamics~\cite{Ayon-Beato:2000mjt}. Given therefore this connection between repulsive-like behavior and alternative gravity theories, one can promote PBHs with repulsive behavior as a novel portal to probe the potential quantum nature of gravity, exploiting as well the associated to these ultralight repulsive PBHs observational signatures~\cite{Flachi:2012nv,Papanikolaou:2020qtd,Domenech:2020ssp,Anantua:2008am,Dong:2015yjs,Ireland:2023avg,Calza:2025whq}. 

Finally, it is important to stress here that early matter dominated (eMD) eras, like the ones driven by PBHs and considered in this Letter, facilitate the growth of early structures. This is expected from the fact that sub-horizon energy density fluctuations during an eMD era grow linearly with the scale factor, i.e. $\frac{\delta \rho}{\rho}\propto a$~\cite{Mukhanov:1990me}, leading potentially to the formation of early structures such as halos or virialised objects. This possibility has been extensively studied in the recent years~\cite{Jedamzik:2010dq, Barenboim:2013gya,Eggemeier:2020zeg,Hidalgo:2022yed,Domenech:2023afs} in association as well with gravitational wave (GW) production~\cite{Dalianis:2020gup,Fernandez:2023ddy,Dalianis:2024kjr}. PBH-driven eMD eras can be accessed as well indirectly through GW probes associated to PBH number density isocurvature perturbations~\cite{Papanikolaou:2020qtd,Domenech:2020ssp,Papanikolaou:2022chm} as well as to Hawking-radiated gravitons~\cite{Anantua:2008am,Dong:2015yjs,Ireland:2023avg}.


{\bf{Acknowledgments}}--
The authors acknowledge enlightening discussions with Andrei Zelnikov, Bernard Carr, M. Sakellariadou and Konstantinos Dimopoulos as well as the contribution of the COST Action CA21136 ``Addressing observational tensions in cosmology with systematics and fundamental physics (CosmoVerse)''. K.F.D. was supported by the PNRR-III-C9-2022–I9 call, with project number 760016/27.01.2023. TP acknowledges the contribution of the LISA Cosmology Working Group and the COST Action ``CA23130 - Bridging high and low energies in search of quantum gravity (BridgeQG)'' as well as the support of the INFN Sezione di Napoli \textit{iniziativa specifica} QGSKY. The author names are shown alphabetically.

\appendix\label{eq:Appendix}
\appendix\label{SwCh}
\section{Swiss cheese matching}\label{swch}

The matching of an exterior homogeneous and isotropic metric to an interior black hole metric 
is made across a spherical 3-surface $\Sigma$ which is at fixed coordinate radius in the cosmological frame but evolving in the black hole frame~\cite{Einstein:1945id}. 
The matching is guaranteed if the first fundamental form (intrinsic metric) and second fundamental form
(extrinsic curvature), calculated in terms of the coordinates on $\Sigma$, are
the same on both sides \cite{Darmois:1927,Stephani:1990, Dyer:2000sn}.

The general cosmological metric can be written in spherical coordinates as
\begin{equation}
ds^2=-dt^2+a^2(t)\left[r^2\left(d\theta^2+\sin^2\theta \,  d\phi^2\right)                       +\frac{dr^2}{1-kr^2}\right]\, ,
\label{eq:FRW}
\end{equation}
where $a(t)$ is the scale factor and $k=0, \pm 1$ is the spatial curvature constant.
 
The Darmois-Israel junction conditions \cite{Eisenhart:1949} allow us to use different coordinate systems on both sides of the
hypersurface. This allows the metric \eqref{eq:FRW} to be joined smoothly to a static and spherically symmetric metric of the form,
\begin{equation}
ds^2=-F(R)\,dT^2
             +R^2\left(d\theta^2+\sin^2\theta d\phi^2\right)
             +\frac{1}{F(R)}\,dR^2\, .
\end{equation}
The first fundamental form is the metric on $\Sigma$ induced by the
spacetime in which it is embedded. This may be written as
\begin{equation}
\gamma_{\alpha\beta}=g_{ij}\frac{\partial x^i}{\partial u^\alpha}
\frac{\partial x^j}{\partial u^\beta}\, , 
\end{equation}
where $u^\alpha=(u^1\equiv u,\, u^2\equiv v,\, u^3\equiv w)$ 
 is the coordinate system
on the hypersurface. Greek indices run over $1,\ldots ,3,$ while Latin indices over
$1,\ldots ,4.$ The second fundamental form \cite{Eisenhart:1949} is defined by
\begin{equation}
K_{\alpha\beta}=n_{i;j}\,\frac{\partial x^i}{\partial u^\alpha}\frac{\partial x^j}{\partial u^\beta}=(\Gamma^p{}_{ij}n_p-n_{i,j})
                      \frac{\partial x^i}{\partial u^\alpha}\frac{\partial x^j}{\partial u^\beta}\, , 
\end{equation}
where $n_a$ is a unit normal to $\Sigma$ and $\Gamma^p{}_{ij}$ are the Christoffel symbols.
We use subscripts $F$ and $S$ to denote quantities associated with the cosmic and
black hole metric, respectively.

The spherical hypersurface $\Sigma$ is given by the function
$f_F(x_F^i)=r-r_\Sigma=0,$ where $r_\Sigma$ is a constant.
The hypersurface coordinates in the FLRW frame ($x_F$) are
$(t=u,\, \theta=v,\, \phi=w,\,r=r_\Sigma )$, while those in
the black hole frame ($x_S$) are
$(T=T_S(u),\,\theta=v,\, \phi=w,\,R=R_S(u))$. Since
$R=R_S(u)$,
$\Sigma$ cannot remain 
constant in the black hole radial coordinate as the Universe expands. This radial distance is also called Schucking radius~\cite{Schucking:1954} and the successful matching shows that this choice of the matching surface is appropriate.
In the following,  $T$ and $R$ mean $T_S$ and $R_S$, respectively. 
The corresponding differential equations derived from the first and second matching conditions provide the expansion rate, the acceleration of the scale factor and a constraint equation between $R$ and $r_\Sigma$. In particular, the first condition, $\gamma_{F\alpha\beta}=\gamma_{S\alpha\beta}$, gives
\begin{equation} \label{trivial}
-1=-F(R)\,\left(\frac{dT}{du}\right)^2
    +F(R)^{-1}\left(\frac{dR}{du}\right)^2\, 
\end{equation}
and 
\begin{equation}
a\,r_{\Sigma}=R,       \label{eq:First FF2}
\end{equation}
while the second matching condition, for the extrinsic curvatures, i.e. $K_{S\alpha\beta}=K_{F\alpha\beta}$ on $\Sigma$, using also Eqs \eqref{trivial} and \eqref{eq:First FF2}, gives
\begin{eqnarray}
\frac{dT}{du}&=&\pm\frac{\sqrt{1-k\,r_\Sigma^2}}{F(R)}\, \label{times} \\
\Big(\frac{dR}{du}\Big)^{2}&=&1\!-\!k r_{\Sigma}^{2}-F(R)\label{gyyu} \\
\frac{d^{2}R}{du^{2}}&=&-\frac{F'(R)}{2}, \label{kevf}
\end{eqnarray}
with a prime denoting differentiation with respect to $R$ and where $F$ and $R$ are evaluated at the Schucking radius $R_S= a\,r_{\Sigma}$. 

For the case of the Schwarzschild metric, i.e. $F(R)=1- 2G_\mathrm{N} m/R$, Eqs \eqref{times}, \eqref{gyyu} and \eqref{kevf} give
\begin{equation}\label{fr1cl}
  H^2=\frac{\dot{a}^2}{a^2}=\frac{2G_\mathrm{N} m}{R^3}-k\frac{r_\Sigma^2}{R^2} = \frac{8\pi G_\mathrm{N}}{3}\rho_{\rm m}- \frac{k}{a^2} \, ,
\end{equation}
where in the last equality we used \Eq{eq:First FF2}. Interestingly enough, one obtains the conventional Friedmann equation
and assume that the matter is in the form of homogeneously distributed black holes, i.e. $\rho_{\rm m} = 3m/ (4\pi R^3)$
Furthermore, \Eq{gyyu}
gives
\begin{equation}\label{fr2cl}
  \ddot{a}=-\frac{G_\mathrm{N} m}{R^2}\frac{1}{r_\Sigma}=-\frac{4\pi G_\mathrm{N}}{3}a\,\rho_m \, ,
\end{equation}
which indicates 
deceleration, as expected. This is the conventional ``Swiss Cheese'' or Einstein-Strauss model that generates a dust Universe, introduced firstly in~\cite{Einstein:1945id}.

One needs to stress here that the matching of a black hole solution with a cosmological metric happens on a mathematical surface with radius equal to the Shcuking radius. This mathematical surface is not static, it expands. If the matching happens inside the horizon, the latter does not cause any problem since there is no violation of any physical law or principle. The matching surface is a mathematical surface. One actually matches two different solutions of the Einstein field equations obtaining after the matching, a configuration being also a solution of the field equations.

A cosmological spacetime filled with many black holes, arises by simply repeating the above process for many disjoint spherical regions in the FLRW background \cite{Einstein:1945id}, \cite{Schucking:1954}, \cite{Carrera:2008pi}. We cut away multiple non-overlapping spheres from the FLRW spacetime and replace each with a spherical black hole region and then matches it using the same junction conditions. Given that the excised spheres do not overlap, each black hole region is matched locally, and the remaining cosmological background still satisfies the gravitational field equations with the same matter content. This repetition is possible because field equations are local: each excised region can be matched independently. Furthermore, the matching is purely geometrical and does not require global symmetries beyond those of each patch. The spheres are typically chosen to be comoving with the cosmological expansion so that the junction surface maintains its shape in the FLRW coordinates. The large-scale homogeneity and isotropy is recovered because the holes can be distributed in a statistically homogeneous and isotropic way. 
Thus, for the case of a population of many black holes, we need to work with energy densities~\cite{Schucking:1954}, \cite{Carrera:2008pi}. Note that $r_{\Sigma}$ is a free parameter in our ``Swiss Cheese" model and it is constant, unlike the matching (Schucking) radial coordinate of the black hole metric $R$. For a more detailed explanation see \cite{Schucking:1954}, \cite{Carrera:2008pi}. 
In the very early Universe, when $a\rightarrow 0$,  one can have very high energy densities $\rho$ larger than the characteristic one $\rho_\mathrm{ch}$ so as to get cosmic acceleration and at the same time non overlapping black holes (cosmic initial conditions concerning $a$ and $r_\Sigma$). For non overlapping we must have for a flat cosmology $N\,r_\Sigma\,a_i<r_{un}\,a_i$, where $N$ is the number of black holes, $a_i$ the scale factor when PBHs form and $r_{un}$ the length size of the Universe at that time. This inequality can always be true since $r_\Sigma$ is free parameter ($a_i$ decouples).
The matching can happen also for horizonless spherical solutions with no contradiction with the Darmois-Israel junction conditions.

\section{The Bardeen metric}\label{Appendix_Bardeen}
The Bardeen spacetime is described by the following line element \cite{Konoplya:2023ahd}
\begin{equation}
\mathrm{d} s^2 = - f(R) \mathrm{d}t^2 + f^{-1}(R) \mathrm{d}R^2 + R^2 \mathrm{d} \Omega ^2\,,    
\end{equation}
where
\begin{equation}
    f(R) = 1 - \frac{2 G_\mathrm{N} m R^2}{(R^2 + l_0^2)^{3/2}}\,.
\end{equation}
This function has a single minimum at $R_{\rm min} = \sqrt{2}l_0$ with $f(R_{\rm min})$ vanishing for $l_0 = 4 G_\mathrm{N} m/(3\sqrt{3})$. From the regularity of the curvature invariants \cite{Ayon-Beato:2000mjt}, it follows that, as long as $|l_0| \leq \frac{4G_\mathrm{N} m}{3\sqrt{3}}$, the above spacetime possesses only coordinate-singularities that describe the existence of event horizons. In greater detail, for the exact equality, the two horizons shrink into one, which corresponds to an extreme black hole, just as in the case of the Reissner-Nordstr\"om solution. For the pure inequality, there exist an inner and an event horizon at the roots of $f(R) = 0$, which we omit to give here for simplicity. The interested reader can check \cite{Ayon-Beato:2000mjt,Borde:1996df} for more details. In general though, the global structure of Bardeen spacetime resembles that of the Reissner-Nordstr\"om one, but instead of a singularity at $R = 0$, $R = 0$ just represents the origin of spherical coordinates.

Following the same procedure as in the Hayward spacetime regarding the matching of the black hole spacetime with an expanding homogeneous and isotropic Universe, the relevant matching conditions \eqref{gyyu} and \eqref{kevf} for the Bardeen spacetime will take the form
\begin{align}
    \left( \frac{\mathrm{d} R}{\mathrm{d}t}\right)^2 &= \frac{2 G_\mathrm{N} m R^2}{\left(R^2+l_0^2\right)^{3/2}}-k r_{\Sigma}^2\,, \\
    \frac{\mathrm{d}^2 R}{\mathrm{d}t^2} & = \frac{G_\mathrm{N} m \left(2 l_0^2 R-R^3\right)}{\left(R^2+l_0^2\right)^{5/2}}\,.
\end{align}
Thus, since $a r_{\Sigma} = R$ the cosmic evolution equations will read as
\begin{gather}
     \frac{\dot{a}^2}{a^2} = \frac{2 G_\mathrm{N} m}{(R^2 + l_0 ^2)^{3/2}} - \frac{k}{a^2}\,,\\
    \frac{\ddot{a}}{a} = \frac{G_\mathrm{N} m ( 2 l_0^2-R^2)}{(R^2 + l_0^2)^{5/2}}\,.
\end{gather}
Using the ``Swiss Cheese" property $\rho = 3m /(4\pi R^3)$ one can express the above equations as
\begin{gather}
    H^2 + \frac{k}{a^2} = \frac{2 G_\mathrm{N}m}{ \left[\left(\frac{3m}{4\pi\rho}\right)^{2/3}+l_0^2\right]^{3/2}}  \,,\\
    \frac{\ddot{a}}{a} = \frac{  G_\mathrm{N}m \left[2 l_0^2-\left(\frac{3m}{4\pi\rho}\right)^{2/3}\right]}{ \left[\left(\frac{3m}{4\pi\rho}\right)^{2/3}+l_0^2\right]^{5/2}}\,.
\end{gather}

One then obtains acceleration if the density is above a characteristic value reading as
\begin{equation}
    \rho_\mathrm {ch}=\frac{3}{\pi}2^{-7/2}m\,l_0^{-3}.
\end{equation}
Therefore, for this Bardeen regular black hole spacetime one can end the early cosmic acceleration phase either due to evaporation or when the black hole density drops below $\rho_\mathrm {ch}$.

\section{The Dymnikova metric}\label{Appendix_Dymnikova}
The Dymnikova black hole is described by the following line element \cite{Dymnikova:1992ux}
\begin{eqnarray}
    \mathrm{d}s^2 = - f(R) \mathrm{d}t^2 + f(R)^{-1}\mathrm{d}R^2 + R^2 \mathrm{d} \Omega ^2\,,
\end{eqnarray}
with $$f(R) = 1 - \frac{\mu}{R}\left( 1- e^{-\frac{R^3}{\alpha \mu}}\right)\,,$$
where $\mu = 2 G_\mathrm{N} m$ and $\alpha = 3/\Lambda$.  This solution coincides with the Schwarzschild black hole for $R \gg (\alpha \mu)^{1/3}$, while for $R \ll (\alpha \mu)^{1/3}$ it behaves like the de Sitter solution. It possesses two event horizons located approximately at 
\begin{equation}
R_+ \simeq 2 G_\mathrm{N}   m \quad \text{and} \quad R_- \simeq \sqrt{\alpha}\,,    
\end{equation} 
where $R_+$ is the external event horizon and $R_-$ is the Cauchy horizon. Both of them are coordinate-singularities and can be removed by an appropriate coordinate transformation.

Applying again the ``Swiss Cheese'' matching conditions to the Dymnikova spacetime one gets similarly the following constraint equations:
\begin{gather}
    \left( \frac{\mathrm{d} R}{\mathrm{d}t}\right)^2 = \frac{\mu}{R}\left(1-e^{-\frac{R^3}{\alpha  \mu }}\right) - k r_{\Sigma}^2 \,, \\
    \frac{\mathrm{d}^2 R}{\mathrm{d}t^2}  = \left(\frac{\mu }{2 R^2}+\frac{3 R}{2 \alpha }\right) e^{-\frac{R^3}{\alpha  \mu }}-\frac{\mu }{2 R^2},
\end{gather}
while the cosmic expansion equations will be recast as
\begin{gather}
     \frac{\dot{a}^2}{a^2} = \frac{\mu}{R^3} \left( 1- e^{-\frac{R^3}{\alpha \mu}}\right) - \frac{k}{a^2}\,,\\
    \frac{\ddot{a}}{a} = -\frac{\mu}{2 R^3} + \left( \frac{3}{2\alpha} + \frac{\mu}{2R^3}\right) e^{-\frac{R^3}{\alpha \mu}}\,.
\end{gather}
Using finally the ``Swiss Cheese'' property $\rho = 3m/(4\pi R^3)$
the above equations are ultimately written as
\begin{gather}
     H^2 + \frac{k}{a^2} = \frac{8 \pi  G_\mathrm{N}}{3}  \rho  \left(1-e^{-\frac{3}{8 \pi G_\mathrm{N}  \alpha  \rho }}\right) \,,\\
    \frac{\ddot{a}}{a} =  - \frac{4\pi G_\mathrm{N}}{3} \rho + \left(\frac{3}{2 \alpha } + \frac{4\pi G_\mathrm{N}}{3} \rho \right) e^{-\frac{3}{8 \pi  \alpha  G_\mathrm{N} \rho }}\,.
    \label{Dumaccel}
\end{gather}
Here, the acceleration holds while 
\begin{equation}
 \rho \gg \frac{\Lambda}{8\pi G_\mathrm{N}}   
\end{equation}
since the exponential in \eqref{Dumaccel} approaches $1$ and so $ \frac{\ddot{a}}{a} \simeq \frac{\Lambda}{2} $. For low values of $\rho$ we get deceleration because the first term in \eqref{Dumaccel} dominates.

\section{Schwarzschild de-Sitter black hole}\label{dSS}

The Schwarzschild-de Sitter (SdS) black hole is a solution to Einstein's field equations in general relativity that describes a spherically symmetric, non-rotating black hole in a homogeneous and isotropic spacetime with a positive cosmological constant \cite{Kottler:1918cxc} being recast as
\begin{eqnarray}
\mathrm{d}s^2&=&-\Big(1-\frac{2G_\mathrm{N} m}{R}-\frac{1}{3}\Lambda R^2\Big)\mathrm{d}T^2 \nonumber \\
&+&\frac{\mathrm{d}R^2}{1-\frac{2G_\mathrm{N} m}{R}-\frac{1}{3}\Lambda R^2}
+R^2\left(\mathrm{d}\theta^2+\sin^2\theta \mathrm{d}\varphi^2\right)\, .
\label{ASFM}
\end{eqnarray}
It is convenient to define $ l^2=\frac{3}{\Lambda}$. Following \cite{Shankaranarayanan:2003ya}, the SdS spacetime has two horizons if $0
< y < 1/27$, where $y \equiv (G_\mathrm{N} m)^2/l^2$. There exist namely the black hole horizon, $R_\mathrm{h}$, and the
cosmological horizon, $R_\mathrm{c}$, given by
$R_\mathrm{h} = \frac{2 G_\mathrm{N}\,m}{\sqrt{3y}} \cos \frac{\pi + \tau}{3} \,$  and
$R_\mathrm{c} = \frac{2 G_\mathrm{N}\,m}{\sqrt{3y}} \cos \frac{\pi - \tau}{3} \,$ respectively, where $\tau = \cos^{-1}(3 \sqrt{3 y}) \, $.
In the limit of $y << 1$, one gets $R_\mathrm{h} \to 2G_\mathrm{N}\,m$ and $R_\mathrm{c} \to l$. The smallest positive root is the black hole horizon, i.e. $R_\mathrm{h} < R_\mathrm{c}$. In the limit $y\to 1/27$, black hole and cosmological horizons coincide and one gets the Nariai spacetime.

Assuming now that PBHs are described by the SdS metric \eqref{ASFM} one can apply the ``Swiss Cheese'' model to derive the cosmic expansion dynamics.
Following the same steps as before, \Eq{gyyu} provides through \Eq{eq:First FF2} the
background cosmic expansion equation being recast as
\begin{equation}
\frac{\dot{a}^2}{a^2}+\frac{k}{a^2}=\frac{2G_\mathrm{N} m}{R^3}
+\frac{\Lambda}{3}\,.
\label{frke}
\end{equation}
In addition, equation (\ref{kevf}) gives
\begin{equation}
\frac{\ddot{a}}{a}=-\frac{G_\mathrm{N}m}{R^3}+\frac{\Lambda}{3}\,,
\label{frjw}
\end{equation}
Using again the ``Swiss Cheese'' property $\rho= 3m/(4\pi R^3)$ Eqs \eqref{frke} and \eqref{frjw} can be written as
\begin{gather}
H^{2}+\frac{k}{a^{2}} = \frac{8\pi G_\mathrm{N}}{3}\rho+\frac{\Lambda}{3},\label{wiiw}\\
\frac{\ddot{a}}{a} = -\frac{4\pi G_\mathrm{N}}{3}\rho+\frac{\Lambda}{3}\,,\label{erjq}
\end{gather}
being the same with them of the $\rm{\Lambda}CDM$ model. However, in our case the physical interpretation is different. We assume that the $\Lambda$ term is the dominant one in order to have an early inflationary phase driven by the cosmic homogeneous soup of SdS PBHs. Thus, we assume that 
\begin{equation}\label{ineq2}
    \Lambda > 4\pi G_\mathrm{N}\,\rho .
\end{equation}
If inequality \eqref{ineq2} holds, then due to Eqs. \eqref{wiiw} and \eqref{erjq}, this entails an endless acceleration. Such an endless acceleration can naturally be terminated through Hawking evaporation. One can exit thus from this early cosmic acceleration phase as soon as PBHs will evaporate considerably to radiation. 

\bibliography{references}

\end{document}